\documentclass[submission,
copyright,
creativecommons,
]{eptcs}
\usepackage{graphicx, amssymb}
\usepackage{amsmath}
\usepackage{amsthm}
\usepackage{mathpartir}
\usepackage{verbatim}
\usepackage{color}
\usepackage{timestamp}
\usepackage{fancyvrb}
\usepackage[normalem]{ulem} 
\usepackage{float}
\usepackage{wrapfig,subfig}
\usepackage[compact]{titlesec}  
\floatstyle{boxed}
\restylefloat{figure}   

\newcommand{\place}{{\tt place}}

\newcommand{\syntaxdef}{\mathrel{::=}}
\newcommand{\ou}{\;\;  \mid \;\; }

 \newtheorem{fact}{Fact}[section]

 \newtheorem{definition}[fact]{Definition}

\newcommand{\these}{\vdash}

\newcommand{\rt}{{\tt r}}
\newcommand{\rd}{{\tt rad}}

\newcommand{\emptyseq}{\emptyset}

\newcommand{\defs}{\texttt{env}}
\newcommand{\FN}{\texttt{FN}}
\newcommand{\FV}{\texttt{FV}}

\newcommand{\fns}{\texttt{fn}}

\newcommand{\parop}{\mid}
\newcommand{\nil}{\boldsymbol {0}}
\def\new#1{\scop{#1}}

\def\pep#1{#1.}
\def\scop#1{\pep{(\nu #1)}}

\newcommand{\mov}{\texttt{mov}}

\newcommand{\dis}[2]{{\tt dis}(#1,#2)}

\newcommand{\delay}{\texttt{delay}}

\newcommand{\x}{x}

\newcommand{\Pexp}{\delta}

\newcommand{\ty}{T}

\newcommand{\ch}{{\tt chan}}
\newcommand{\val}{v}
\newcommand{\tuple}[1]{{#1}}
\newcommand{\chType}[1]{\ch\{#1\}}
\newcommand{\tupleType}[2]{#1_1\ast\cdots\ast #1_{#2}}

\newcommand{\floatType}{{\tt fl}}
\newcommand{\topType}{\top}
\newcommand{\proj}[2]{#1.#2}
\newcommand{\tyEnv}{\Gamma}
\newcommand{\HSep}{\hbox to \textwidth{\bf\hrulefill}}
\newcommand{\emptyTuple}{(\,)}
\newcommand{\typeOf}{{\tt typeOf}}
\newcommand{\const}{{\tt c}}

\newcommand{\WTExp}[3]{#1\vdash #2 : #3}
\newcommand{\WTProc}[2]{#1\vdash #2\ \diamond}

\newcommand{\Po}[3]{\langle#1,#2,#3\rangle}
\newcommand{\this}{{\tt this}}
\newcommand{\Points}{{\it Ps}}
\newcommand{\Scale}{{\it Sc}}

\newcommand{\op}{{\tt op}}
\newcommand{\fst}{{\tt fst}}
\newcommand{\snd}{{\tt snd}}
\newcommand{\conv}{\Downarrow}
\newcommand{\ev}[2]{#1\conv#2}

\newcommand{\OK}{{\rm SC}}

\newcommand{\transl}{{\tt rand}}

\newcommand{\redDet}[1]{\rightarrowtail}
\newcommand{\redStoc}[1]{\dashrightarrow}

\newcommand{\labRed}[1]{\mathrel{{\xrightarrow{#1}}}}
\newcommand{\labPreRed}[1]{{\stackrel{#1}{\hookrightarrow}}}
\newcommand{\preRedSt}[1]{\labPreRed{\texttt{r}}}
\newcommand{\preRedStD}[1]{\labPreRed{#1}}
\newcommand{\preRedMv}[1]{\labPreRed{\texttt{mv}}}

\newcommand{\g}{\xi,\omega, \sigma}

\newcommand{\locP}[3]{\{#1\}_{#3}}
\newcommand{\labP}[1]{\ensuremath{\scriptstyle{\textsc{(#1)}}}}

\definecolor{brown}{rgb}{0.85,.66,0}

\newcommand{\NamedRule}[3]{\scriptstyle{\textsc{(#1)}}\
\displaystyle                  
\frac{#2}{#3}\           
}

\newcommand{\la}[1]{{\scriptstyle{\textsc{(#1)}}}}

\newcommand{\ThreePi} {$3\pi$}
\newcommand{\LBS} {BioScape$^L$}
\newcommand{\origin} {\maltese}

\newif\ifsubmit
\submittrue
\ifsubmit
\newcommand{\AC}[1]{#1}
\newcommand{\ACComm}[1]{}

\newcommand{\MDComm}[1]{}

\newcommand{\PGComm}[1]{}

\newcommand{\ATComm}[1]{}

\newcommand{\mdc}[1]{}

\newcommand{\acc}[1]{}

\newcommand{\atc}[1]{}
\else
\newcommand{\AC}[1]{\textcolor{blue}{#1}}
\newcommand{\ACComm}[1]{{\scriptsize \textcolor{blue}{[Adriana{:} #1]}}}

\newcommand{\MDComm}[1]{{\scriptsize \textcolor{red}{[Mariangiola{:} #1]}}}

\newcommand{\PGComm}[1]{{\scriptsize \textcolor{magenta}{[Paola{:} #1 ]}}}

\newcommand{\ATComm}[1]{{\scriptsize \textcolor{brown}{[Angelo{:} #1 ]}}}

\newcommand{\mdc}[1]{\textcolor{red}{[Mariangiola{:} #1]}}

\newcommand{\acc}[1]{\textcolor{blue}{[Adriana{:} #1]}}

\newcommand{\atc}[1]{\textcolor{brown}{[Angelo{:} #1]}}

\fi

\newcommand{\RC} {Restricted Choice}
\newcommand{\rc} {restricted choice}

\makeatletter
 \let\@copyrightspace\relax
 \makeatother

\title{A Calculus of Located Entities} 

\author{ 
Adriana Compagnoni\institute{Department of Computer Science \\
Stevens Institute of Technology\\
New Jersey, USA\\
Adriana.Compagnoni@stevens.edu} 
\and 
 Paola Giannini\thanks{Partly funded by ``Progetto MIUR PRIN CINA Prot. 2010LHT4KM''.} 
\institute{Computer Science Institute\\
DISIT, Univ. Piemonte Orientale\\
Alessandria, Italy\\
giannini@di.unipmn.it} 
\and 
Catherine Kim
\institute{Department of Computer Science \\
Stevens Institute of Technology\\
New Jersey, USA\\
ckim@stevens.edu} 
\and
Matthew Milideo
\institute{Department of Computer Science \\
Stevens Institute of Technology\\
New Jersey, USA\\
mmiledeo@stevens.edu} 
\and 
 Vishakha Sharma
\institute{Department of Computer Science \\
Stevens Institute of Technology\\
New Jersey, USA\\
vsharma1@stevens.edu} }

\def\titlerunning{{A Calculus of Located Entities}}

\def\authorrunning{Compagnoni, Giannini, Kim, Milideo, Sharma}
\begin{document}
\maketitle

\begin{abstract}
   We define \LBS, a stochastic pi-calculus in 3D-space. A novel aspect
  of \LBS{} is that entities have \textbf{programmable locations}. The
  programmer can specify a particular location where to place an
  entity, or a location relative to the current location of the
  entity.  The motivation for the extension comes from the need to
  describe the evolution of populations of biochemical species in
  space, while keeping a sufficiently high level description, so that
  phenomena like diffusion, collision, and confinement can remain part
  of the semantics of the calculus.  Combined with the random
  diffusion movement inherited from BioScape, programmable locations
  allow us to capture the assemblies of configurations of polymers,
  oligomers, and complexes such as microtubules or actin filaments.

  Further new aspects of \LBS{} include \textbf{random translation}
  and \textbf{scaling}. Random translation is instrumental in
  describing the location of new entities relative to the old
  ones. For example, when a cell secretes a hydronium ion, the ion
  should be placed at a given distance from the originating cell, but
  in a random direction. Additionally, scaling allows us to
  capture at a high level events such as division and growth; for
  example, daughter cells after mitosis have half the size of the
  mother cell.  
\end{abstract}

\section{Introduction}

{
Our earlier work on BioScape\cite{BioScape:ENTCS} was motivated by the
need to visualize the evolution of species in 3D space. The simulator
of BioScape randomly places initial distributions of entities within
specified confinement areas.\footnote{Models of biological and
  biomedical applications using BioScape can be
found in Compagnoni's website.}
However, while BioScape naturally captures a large
family of wet-lab experiments, it does not have the ability to
describe the assembly of entities into compound structures such as
dimers, polymers, oligomers, etc. In order to describe the composition
of such structures from smaller components, we introduce a new
calculus where an entity's location can be programmed. The long-term goal of our research
program is to create programming platforms to describe complex 3D landscapes, where agents
interact with the environment. Applications of such modeling platforms
include simulating intracellular viral traffic, and designing multifunctional antibacterial surfaces
that prevent or minimize infection while 
 maximizing  tissue growth.}
 
{
In this paper we define \LBS, a stochastic $\pi$-calculus in 3D-space
with programmable locations.
It builds on BioScape\cite{BioScape:ENTCS} by adding three new
features: programmable entity's location, random translation and
scaling. As we just mentioned, programmable locations allow the
programmer to specify the location of new entities, either by
describing an absolute location in the global frame, or by specifying
a location relative to the current location of the generating
entity. Random translation lets the programmer describe a distance
from the original position where to place the new entity without
specifying an absolute or relative location. For example a random
translation from point \texttt{p} of 1cm  will place the new entity's
barycentre somewhere on  the 1cm radius sphere
around \texttt{p}. Finally, scaling enables the creation of new
entities whose shape is obtained by resizing the shape of  the
original entity. The key aspect of all three extensions is their high
level nature.  The placement of new objects in space needs to account
for confinement and collision, which in \LBS{} are part of the 
semantics of the calculus, unlike in low level calculi, where they are a burden
to the programmer.
}

{ 
As we observed before, dynamic spatial arrangement of
  components is useful in representing assembly of polymers such as
  actin filaments and cytoskeletal microtubules.  Microtubules are
  part of the cytoskeleton of eukaryotic cells, and form roads on
  which organelles ride on their way to the cell nucleus.
  Microtubules are hollow and formed with dimers of $\alpha$ and
  $\beta$ tubulin. They are anchored to a starting point around the
  Microtubules Organizing Center, and while the starting point is
  fixed, microtubules grow and shrink from the end piece. We now motivate
  the programmable entity's location feature, by implementing a simplified
  model of microtubules polymerization in \LBS.  Random translation
  and scaling are introduced later in Section \ref{sec:RandomScale}.
}

\begin{figure}
\begin{minipage}[c]{0.6\textwidth}
{
\footnotesize
\begin{verbatim}
val Cytosol:space =  cuboid(50.0,50.0,30.0) @ <1.0,2.0,24.0>
val step = 0.0,stepP = 0.1, r = 0.0, rP= 0.2
new MTConstruction@0.116,rP:ch(ch(),fl*fl*fl) 

let MTPart()@Cytosol,stepP,sphere(1.0)=( new y@0.27,r:ch()
do ?MTConstruction(x,u); MTLeft(x)_glue(this,u)
or !MTConstruction(y,this); MTRight(y)_this  
or mov.MTPart()_this )

and MTRight(rht:chan())@Cytosol,step,sphere(1.0) =
do delay@1.0; MTRight(rht)_this 
or ?rht; MTPart()_this

and MTLeft(lft:chan())@Cytosol,step,sphere(1.0) = 
( new z@0.27,r:ch() 
do delay@1.0; MTLeft(lft)_this 
or !MTConstruction(z,this); MTMiddle(lft,z)_this  
or !lft; MTPart()_this 
or ?lft;MTPart()_this )

and MTMiddle(rht1:chan(),lft1:chan())@Cytosol,
     step,sphere(1.0) = 
do delay@1.0; MTMiddle(rht1,lft1)
or !lft1;MTLeft(rht1)_this

run (MTPart()_p1 | MTPart()_p2 |...| MTPart()_pN )
\end{verbatim}
}
\end{minipage}
\begin{minipage}[c]{0.69\textwidth}
\includegraphics[width=\textwidth]{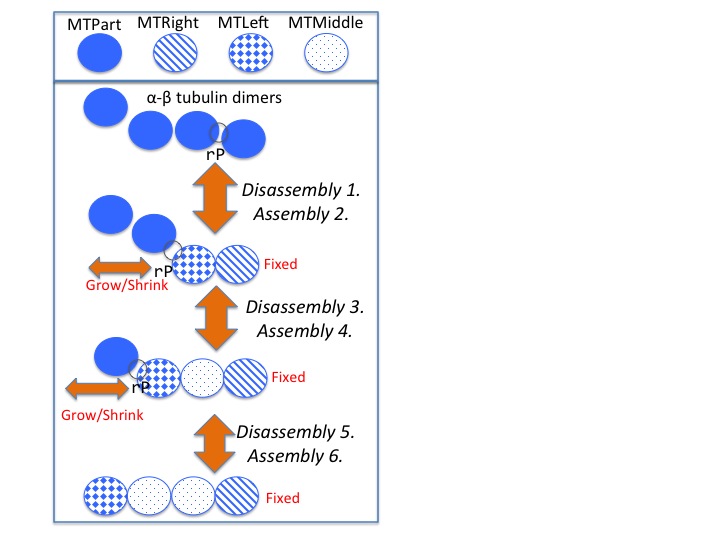}
\end{minipage}
\caption{Microtubules polymerization}
\label{fig:MicroBioScapeSpatial}
\end{figure}

\subsection*{A motivating example} 
{For our next example, microtubules polymerization, consider Fig.\
\ref{fig:MicroBioScapeSpatial}, containing the \LBS{} code as well as
a graphical representation of the evolution of the system. Microtubules are dynamic tubulin polymers; although they 
are formed with dimers of $\alpha$ and $\beta$ tubulin, we simplify 
their structure in our example, and consider them as assembled starting from parts, }
\verb!MTPart!, where a part is an $\alpha$-$\beta$ tubulin dimer. \verb!MTPart!s
are scattered in the Cytosol. 
Microtubules have a start piece \verb!MTRight! and an end piece
\verb!MTLeft!. Between the start
(right)  and the end (left) pieces there can be any number 
of \verb!MTMiddle! pieces. While the start piece is fixed, microtubules grow and
shrink  from the end piece. In order to grow, a new \verb+MTPart+
becomes the new \verb!MTLeft!, and the old \verb!MTLeft! becomes an
\verb!MTMiddle!. Similarly the end piece can disassemble making the
last  \verb!MTMiddle!  the new \verb!MTLeft!, and making the old
\verb!MTLeft! a free \verb!MTPart!.  
The construction is done using
private channels, similar to the process modeling of actin polymerization of \cite{CardelliActin09}, so that only adjacent pieces share channels.
In this model, we assume that \verb!MTLeft!,
\verb!MTMiddle!, and  \verb!MTRight! do not move, unless they become  a free \verb!MTPart!.

We assume an initial concentration of \verb+N MTPart+'s   placed
in the Cytosol, implemented with a parallel composition of \verb+N+
copies of \verb+MTPart+ with barycentres \verb+p1+, $\cdots$,
\verb+pN+ in the \texttt{run} command at the end of the program.

{The first line of code defines the space within which all the entities are
  enclosed. It is a cuboid whose bottom left vertex is the point
  \texttt{(1.0,2.0,24.0)}. The second line defines four floating point constants
which will be used to specify the {\it step of the diffusion rate} of the entities, and the {\em radius of the channels}.
The diffusion rates are: \texttt{step=0.0} for the components of
  the microtubules, i.e., \texttt{MTLeft},  \texttt{MRight}, and
  \texttt{MTMiddle}, since we assume that they do not move, and 
 \texttt{stepP=0.1} for \texttt{MTPart}s, which are subject to brownian motion.
The radius of a channel is the maximum distance between  two  entities
synchronizing on that channel.
Communications between entities forming  microtubules requires radius
\texttt{r=0.0}, specifying that communication can only happen upon contact.
Instead, the radius \texttt{rP} specifies that for two  entities to
synchronize on  channel   \texttt{MTConstruction}, {their closest
points} must be at most \texttt{0.2} units apart. }

The expression \texttt{new MTConstruction@0.116,rP:ch(ch(),fl*fl*fl)} declares channel \\
 \texttt{MTConstruction}, with \textit{stochastic rate} 0.116, and radius
 \texttt{rP}. The stochastic rate is used by the simulation algorithm
 to determine the probability and the reaction time for
 synchronization on the channel.  The \textit{type}
 \texttt{ch(ch(),fl*fl*fl)}  declares \texttt{MTConstruction}, as a
 channel {on which the data exchanged are pairs}
 whose first component is another channel and the
 second component \AC{is} a triple of floating-point numbers.

In the rest of the program \texttt{MTPart}, \texttt{MTRight},  \texttt{MTLeft}, and
  \texttt{MTMiddle} are defined. Each definition has four
  components. Consider the case of \texttt{MTPart}, the
  \texttt{Cytosol} is the confinement area, where instances of
  \texttt{MTPart} can be located; \texttt{stepP} is the diffusion rate
  of an \texttt{MTPart}, \texttt{sphere(1.0)} is its shape, and the
  rest is a process describing the behavior of \texttt{MTPart}.

An \texttt{MTPart} can  either synchronize with another
\texttt{MTPart} and  become \texttt{MTRight} and
\texttt{MTLeft} respectively. It can also synchronize with an
\texttt{MTLeft}, or move.

In more detail, \AC{for each instance of} \verb!MTPart!, a new private
channel is created with \verb!new y@0.27,r!, where \verb!y! is the
name of the channel. The stochastic reaction rate of the channel is
\verb+0.27+, and the channel radius is \verb+r+.  \verb!MTPart! can
either do an input on channel \verb!MTConstruction!,
\verb!?MTConstruction(x)!, or an output on the same channel,
\verb+!MTConstruction(y)+.

Consider \verb!MTPart()_p1 | MTPart()_p2!,  representing 
\verb!MTPart!'s at locations \verb!p1! and \verb!p2! respectively. If the 
closest points of the two parts are closer that \verb!rP!, there can be a
synchronization on channel \verb!MTConstruction!. 
The entity \verb!MTPart()_p1! sends on channel \verb!MTConstruction!
the private channel
name \verb!y! and the position \verb!p1!, and it becomes \verb!MTRight(y)_p1!, whereas \verb!MTPart()_p2! receives
\verb!y!, and \verb!p1!, on channel \verb!MTConstruction!, binds \verb!y! to \verb!x! and
\verb!u! to \verb!p1!, and it
becomes \verb!MTLeft(y)_p3!. Point \verb!p3!, the result of \verb!glue(p2,p1)!, is such that \verb!MTLeft(y)_p3! and \verb!MTRight(y)_p1!
are in contact with each other. \verb!MTLeft(y)_p3! shares 
the private channel \verb!y! with \verb!MTRight(y)_p1!. 
This evolution is shown in the picture at
Fig.~\ref{fig:MicroBioScapeSpatial}, by Assembly 2.
Note that,  $\this$ denotes the barycentre
of the \verb!MTPart! from which \verb!MTRight! or \verb!MTLeft! evolve. 
The metavariable $\this$ is an abstract reference to  the runtime
position of the generating entity;  $\this$ is similar to the origin, \origin, of
\ThreePi \cite{DBLP:journals/tcs/CardelliG12}. 
The position of an entity can be the result of an operation such as
the sum of points or scalar product derived from the location of the
originating entity (\texttt{this}).

The entity  \verb!MTPart()_p1! can perform the move action, in which case a new point 
\verb!p4! placed randomly at distance \verb!stepP! from \verb!p1! is generated,
and  \verb!MTPart()_p1! evolves into a new \verb!MTPart! located at \verb!p4!.

The entity  \verb!MTRight! can remain an \verb!MTRight!
with a delay prefix, or it can do an input action with the adjacent \verb!MTLeft! with which it shares
the  channel \verb!rht! and evolve into a \verb!MTPart! placed
in its original position (\texttt{this}). This corresponds to the final disassembling of the
microtubule, shown in the picture in
Fig.~\ref{fig:MicroBioScapeSpatial}, by Disassembly 1.
Notice that, in this case, there is no information
sent on  channel \verb!rht!.

The entity \verb!MTLeft! {has a parameter \texttt{lft}, which is a 
channel private to \texttt{MTLeft} and the adjacent \texttt{MTRight}
or \texttt{MTMiddle}. \texttt{MTLeft} }
has four alternative behaviors. It can remain
an \verb!MTLeft! with a delay prefix (first line of the
definition). It can interact with a \verb!MTPart!, by synchronizing on
channel \verb!MTConstruction!, and evolve into a \verb!MTMiddle!  with which it shares
the private channel \verb!z! for interactions, and to which it passes
the {private
channel \texttt{lft}, shared with  adjacent \texttt{MTMiddle}
or \texttt{MTRight}}. In other words, 
\verb!MTLeft(y)_p3 | MTPart()_p4! \AC{becomes}
\verb!MTMiddle(y,z)_p3 | MTLeft(z)_p5!, where \verb!p5! is
\verb!glue(p4,p3)!; see Assembly 4 and 6 in 
Fig.~\ref{fig:MicroBioScapeSpatial}. {In Assembly 4, the channel \texttt{y} is shared 
with the adjacent \texttt{MTRight}, whereas in Assembly 6, it is shared 
with the adjacent \texttt{MTMiddle}. \texttt{MTLeft}} can also interact with a
\verb!MTRight!, by synchronizing on their private channel and
disassemble; see Disassembly 1 in
  Fig.~\ref{fig:MicroBioScapeSpatial}. {Finally,} \texttt{MTLeft} can interact with a
\verb!MTMiddle!  on their private channel, and disassemble; see
Disassembly 5 and 3 in 
Fig.~\ref{fig:MicroBioScapeSpatial}.
For example, consider \verb!MTMiddle(y,z)_p3 | MTLeft(z)_p5!, 
the synchronization on private channel \verb!z!  makes  \verb!MTMiddle(y,z)_p3! evolve into \verb+MTLeft(y)_p3+.  {Alternatively}, with
the same synchronization, \verb!MTLeft(z)_p5! evolves into \verb+MTPart()_p5+, becoming a free part.

{The entity \texttt{MTMiddle} can remain an \texttt{MTMiddle}
with a delay prefix, or it can synchronize with the adjacent \texttt{MTLeft}. As 
previously described, \texttt{MTMiddle(y,z)\_p3} evolves into \texttt{MTLeft(y)\_p3},
which, in Disassembly 5, shares the channel \texttt{y} with an \texttt{MTMiddle}, whereas in 
Disassembly 3, it shares the channel \texttt{y} with the final \texttt{MTRight}.
}

\section{\LBS: Syntax}\label{sec:BioScapeL}

\begin{figure}[t]
  { \begin{align*}
  P,Q & \syntaxdef  \nil & \text{Empty Process}\\
  &\ou X(\Pexp)_{\Pexp} &\text{Located Entity Instance}\\
  &\ou P \parop Q& \text{Parallel Composition}\\
  & \ou \new{a@{\Pexp,\Pexp}:\chType{\ty}}P &\text{Restriction}\\
  M
  &\syntaxdef   \pi.P \; [+\; M] & \text{Choice of Prefixed Process}\\
  \pi &  \syntaxdef  \delay @{\Pexp} & \text{Delay} \\
  &  \ou !u(\Pexp) & \text{Output} \\
  &  \ou ?u(\x)  & \text{Input}\\
  &\ou  \mov &\text{Move}\\
  N
  &\syntaxdef   {M \; \ou \new {a@{\Pexp,\Pexp}:\chType{\ty}} N} & \text{\RC}\\
    u & \syntaxdef a\ou b\ou\cdots \ou x\ou y\ou \cdots  & \text{Identifiers}\\
    \Pexp &\syntaxdef u\ou \const \ou\this\ou \tuple{\Pexp_1,\ldots,\Pexp_n}\ou\emptyTuple\ou\proj{\Pexp}{i}\ou\op(\Pexp)& \text{Expressions}\\ 
     \val & \syntaxdef a\ou b\ou\cdots\ou \const \ou\emptyTuple\ou\tuple{\val_1,\ldots,\val_n}  & \text{Expression Values}\\
  \ty &\syntaxdef \chType{\ty}\ou\floatType\ou\tupleType{\ty}{n}\ou\topType & \text{Expression Types}\\
  D & \syntaxdef \emptyset \ou D,X(\x:\ty) = {N}^{\g}\quad\FV(M)\subseteq\overline{x}& \text{Entity Definitions}  \\
  E & \syntaxdef \emptyset \ou E, {a@\rt,\rd:\chType{\ty}} & \text{Channel Declarations}\\
  \tyEnv & \syntaxdef \emptyset \ou \tyEnv,X{:}\ty\ou \tyEnv,u{:}\ty & \text{Type Environment}
\end{align*}}
\caption{Syntax of \LBS}
 \label{fig.syntax}
\end{figure} 
{The abstract syntax of  {\LBS} extends that of BioScape
\cite{PBioScape:Mecbic}, and it appears  in Fig.~\ref{fig.syntax}. 
We assume a set of {\em channel names}, denoted by
$a$, $b$, and a set of {\em variables}, denoted by $x$,
$y$, and  the metavariable $\const$ for {\em real numbers}. 
We will also use $\rt$ for the stochastic rate, and $\rd$ to
specifying the radius of channels, both $\rt$, and $\rd$,  are real numbers. 
{\em Points}, denoted by the metavariable $p$, are triples $(\const_1,\const_2,\const_3)$
of real numbers. }

Expressions $\delta$ may be channel names, variables, real numbers, the metavariable $\this$, 
tuples of expressions, including the empty tuple $\emptyTuple$, tuple
selection $\proj{\Pexp}{i}$, and operators applied to expressions $\op(\Pexp)$.
The metavariable $\this$ denotes the barycentre of the entity instance {in which
the expression is evaluated.}
Expression values,  are either channel names, real numbers, or tuples
of value.
The \LBS\ types characterizing these values are:  
 channel types, $\chType{\ty}$, specifying the type $\ty$ of the values sent on them;
the type of real numbers, $\floatType$;
the type of tuples, $\tupleType{\ty}{n}$, specifying the types $\ty_i$  of its components, and
 $\topType$, which is the type of the empty tuple. { Channels only used
 for synchronization, such as \texttt{lft} in
 Fig.~\ref{fig:MicroBioScapeSpatial} have type $\chType{\topType}$.}

The empty process is $\nil$.
By $X(\Pexp)_{\Pexp'}$ we denote an instance of the entity defined by
$X$, \AC{with actual parameter $\Pexp$ and positions $\Pexp'$.}
The process $ P \parop Q$ is the parallel composition
of processes $ P$ and  $Q$.
\AC{
 The process
$\new{a@\Pexp,\Pexp':\chType{\ty}}P$ defines the channel name $a$ with
stochastic rate $\Pexp$,  radius $\Pexp'$, and type $\chType{\ty}$ in
process $P$.
As  mentioned before, the radius is the maximum distance between
entities in order to communicate through channel $a$,  the
reaction rate determines how long it takes for two entities to react given that they are close enough
to communicate, and $\chType{\ty}$ states that $a$ is a channel for
communicating values of  type $\ty$.}

The \textit{heterogeneous} choice is denoted by $M$,   where $\pi.P \; [+\;
M]$ means $\pi.P \ou \pi.P \;+\;
M$.    Choices may
  have reaction branches and movement branches.  The reaction branches
  are probabilistic (stochastic), since reactions are subject
  to kinetic reaction rates, while the movement branches are
  non-deterministic, since {diffusion} is always
  enabled. 
%
%
The prefix $\pi$ denotes the action that the process
  $\pi.P$ can perform. The prefix $\delay @\Pexp$ is a
  spontaneous and unilateral reaction of a single process, where
  $\Pexp$ is  
the stochastic
  rate of the reaction. The prefix $!u(\Pexp)$ denotes the output of
  the value of $\Pexp$ on channel $u$,
  and the prefix $?u(x)$ denotes input on channel $u$ with bound
  variable $x$.  
  The prefix $\mov$ denotes the movement of processes in space according to their
  diffusion rate $\omega$.  We use standard syntactic
  abbreviations such as $\pi$ for $\pi.0$.
%
{The \rc{}, denoted by $N$, is a  choice of  prefixed processes $M$ with top level local
channel definitions.}

We denote by $D$ a global list of entity definitions.
The clause $X(\x:\ty)= N^{\g}$
defines entity $X$ with formal parameter $\x$ of type $\ty$ to be 
the \rc{} $N$ with geometry
$\g$, specifying a movement space $\xi$, a step
$\omega$, and a shape $\sigma$. 
The \rc{} $N$ describes the behavior of $X$ with a choice of prefixed processes
$M$, and the set of channels private to the entity $X$.
The movement space $\xi$ is a { 3D area where instances of $X$ are
  allowed to be located.
The step $\omega \in
\mathbb{R}_{\geq0}$, is the distance that $X$ can {move in a unit of time},
and it corresponds to the diffusion rate of $X$;
$\sigma$ is the three-dimensional shape (sphere, cube, etc.) of $X$, having a barycentre.
The movement space for the empty process $\nil$
is everywhere, the global space, and its movement step is 0.
Each entity variable $X$ can be defined at most once  in $D$, and the {\em free variables of
$N$}, must be a subset of the variables $\overline{x}$. 
We also write $X(x)= (\pi.\pi'.P)^{\g}$ as short for
$X(x)= (\pi.Y(x))^{\g}$ and $Y(x)=
(\pi'.P)^{\g}$.  

{\em Free variables}, $\FV$, and {\em free channel names}, $\FN$, of processes and choices 
can be defined in the usual way. The input prefix $?u(x)$, and the restriction $\nu a@\_$
are binders, and define the scope of the variable $x$, and the channel name $a$ respectively.

$E$ ranges over environments of channel name
declarations. $a@\rt,\rd:\chType{\ty}$ defines  channel name $a$ with
rate $\rt$, {radius} $\rd$ and type $\chType{\ty}$. The {\em domain of} $E$ is the set
of channel names declared in $E$, and channel names are declared at most once
in $E$.

$\tyEnv$ ranges over type environments, which map entity names $X$ with the type of the parameter of the entity, channel names $a$ with channel types, and variables with their type. 

In the concrete syntax of the example in
  Fig.~\ref{fig:MicroBioScapeSpatial}, we used \texttt{new} instead of
  $\nu$; \texttt{do-or} instead of ${+}$,  and
\texttt{!a}, \texttt{?a}, and \texttt{chan()}   
instead of 
\texttt{!a()},   \texttt{?a()}, and 
  $\chType{\topType}$,
when no value is exchanged, 

\section{\LBS: Semantics}
We now introduce the {static and dynamic semantics of \LBS. In Fig.~\ref{fig.typing} we define the well formed processes and
definitions, and in Fig.~\ref{fig.opSemExp},~\ref{fig.pi.struct}, and ~\ref{fig.rednorate} the operational semantics of \LBS. }

 In
Fig.~\ref{fig.typing} we define the rules for the judgements:
\begin{itemize}
\item $\WTExp{\tyEnv}{\Pexp}{\ty}$, meaning, {\em in the type environment $\tyEnv$,  the expressions $\Pexp$ has type $\ty$};
\item $\WTProc{\tyEnv}{R}$, meaning, {\em in
    the type environment $\tyEnv$, $R$ is well formed}, where {$R$ is
  either a process $P$, a choice $M$ or a \rc{} $N$,} and
\item $\WTProc{\tyEnv}{D}$, meaning, {\em in the type environment $\tyEnv$, the list of definitions $D$ is well formed}.
\end{itemize}

\begin{figure}[t]
{
    \begin{mathpar}
       \NamedRule{Ty.id}{u{:}\ty\in\tyEnv}
      {\WTExp{\tyEnv}{u}{\ty} }    
      \and 
        \NamedRule{Ty.const}{}
      {\WTExp{\tyEnv}{\const}{\floatType} }    
      \and 
        \NamedRule{Ty.this}{}
      {\WTExp{\tyEnv}{\this}{\floatType\ast\floatType\ast\floatType} }
      \and  
             \NamedRule{Ty.tuple}{\WTExp{\tyEnv}{\Pexp_i}{\ty_i}\quad (1\leq i\leq n)}
      {\WTExp{\tyEnv}{\tuple{\Pexp_1,\ldots,\Pexp_n}}{\tupleType{\ty}{n}} }
    \and
          \NamedRule{Ty.empty}{}
      {\WTExp{\tyEnv}{\emptyTuple}{\topType} }
            \and  
             \NamedRule{Ty.sel}{\WTExp{\tyEnv}{\Pexp}{\tupleType{\ty}{n}}\quad (1\leq i\leq n)}
      {\WTExp{\tyEnv}{\proj{\Pexp}{i}}{\ty_i} }
        \and  
             \NamedRule{Ty.op}{\typeOf(\op)=(\ty_1,\ty_2)\quad\WTExp{\tyEnv}{\Pexp}{\ty_1}}
      {\WTExp{\tyEnv}{\op(\Pexp)}{\ty_2} }
        \end{mathpar} }
\HSep     
{
    \begin{mathpar}   \NamedRule{Ty.nil}{}{\WTProc{\tyEnv} {\nil}}    
\and      
    \NamedRule{Ty.inst}{X{:}\ty\in\tyEnv\quad\WTExp{\tyEnv}{\Pexp}{\ty}\quad\WTExp{\tyEnv}{\Pexp'}{\floatType\ast\floatType\ast\floatType} }
    {\WTProc{\tyEnv} { X(\Pexp)_{\Pexp'}}}    
\and      
    \NamedRule{Ty.par}{\WTProc{\tyEnv} {P}\quad\WTProc{\tyEnv} {Q}}{\WTProc{\tyEnv} {P \parop Q}}    
\and      
    \NamedRule{Ty.restr}{\WTProc{\tyEnv,a{:}\chType{\ty}} {R}\quad\WTExp{\tyEnv}{\Pexp}{\floatType}\quad \WTExp{\tyEnv}{\Pexp'}{\floatType}}{\WTProc{\tyEnv} {\new{a@\Pexp,\Pexp':\chType{\ty}}R}}    
\and      
    \NamedRule{Ty.out}{u{:}\chType{\ty}\in\tyEnv\quad\WTExp{\tyEnv}{\Pexp}{\ty}\quad\WTProc{\tyEnv} {P}} 
    {\WTProc{\tyEnv}{!u(\Pexp).P} 
    } 
 \and      
    \NamedRule{Ty.in}{u{:}{\chType{\ty}}\in\tyEnv\quad\WTProc{\tyEnv,\x{:}\ty} {P}} 
    {\WTProc{\tyEnv}{?u(x).P} 
    } 
\and      
    \NamedRule{Ty.pref}{\WTProc{\tyEnv} {P}} 
    {\begin{array}{l}
    \WTProc{\tyEnv} {\mov.P}
    \end{array} 
    }
    \and      
    \NamedRule{Ty.pref}{\WTProc{\tyEnv} {P}\quad\WTExp{\tyEnv}{\Pexp}{\floatType}} 
    {\begin{array}{l}
    \WTProc{\tyEnv} {\delay @\Pexp.P}
    \end{array} 
    }
\and      
    \NamedRule{Ty.choice}{\WTProc{\tyEnv} {M}\quad\WTProc{\tyEnv} {M'}}{\WTProc{\tyEnv} {M\;+\;M'}}    
\and      
    \NamedRule{Ty.defs}{\WTProc{\tyEnv} {D}\quad\WTProc{\tyEnv,\x{:}\ty} {N}}   
    {\WTProc{\tyEnv} {D,X(\AC{\x:\ty}) = N^{\g}}}    
    \end{mathpar}
    
    }
 
\caption{Well typed expressions, processes, and definitions}
 \label{fig.typing}
\end{figure}
To define the type expressions, we assume a function $\typeOf$ such
that $\typeOf(\op)=(\ty_1,\ty_2)$ means that the operator $\op$ takes
a parameter of type $\ty_1$ and returns a value of type $\ty_2$.  
{The rules for expressions are standard; notice that the type of
  $\this$ in rule $\labP{Ty.this}$ is a triple of floating-points
  representing 3D coordinates.  An entity instance $X(\Pexp)_{\Pexp'}$
  is well formed (rule $\labP{Ty.inst}$), if the actual parameter
  $\Pexp$ has the type associated with $X$ in the type environment,
  and if $\Pexp'$ has the type of a 3D point. In rules $\labP{Ty.out}$
  and $\labP{Ty.in}$ the channel identifier $u$ must have a channel
  type. }


\begin{definition}[{\LBS\ Program, Initial Process, and Initial Configuration}]
  \begin{itemize}
  \item A \LBS{} {\em program} is a triple $(D, E, P)$ such that $D$ is a
    collection of entity declarations, $E$ is a collection of channel
    declarations, and $P$ is a parallel composition of entity
    instances.
  \item  We call $P$ the {\em initial process}.
  \item  We call $E\these P$ the {\em initial configuration} of program  $(D, E, P)$.
  \end{itemize}
\end{definition}
For the example of Fig.~\ref{fig:MicroBioScapeSpatial}, the
initial configuration {is $E\these P$, where $P$ is the argument of the \texttt{run} command:
\[
P {\tt = MTPart()\_p1\ |\ MTPart()\_p2\ |\ ...\ |\ MTPart()\_pN},
\mbox{ and }
\] 
\[
E {\tt  = MTConstruction@0.116,0.2:\chType{\chType{\topType}\ast(\floatType\ast\floatType\ast\floatType)}}
\] 
}
The {\em type environment corresponding to channel declarations or
   entity definitions} $\defs$ is defined as follows, where the
notation $\nu_i$, is an abbreviation for $\nu a_i@\rt_i,\rd_i:\chType{\ty_i}$.
 \begin{definition}[Type Environment]
   \begin{itemize}
   \item $\defs(\emptyset)=\emptyseq$
   \item $\defs(E,
     {a@\rt,\rd:\chType{\ty}})=a{:}\chType{\ty},\defs(E)$
   \item $\defs(D,X(\x:\ty) =
     N^{\g})=X{:}{\ty},\defs(D)$
   \end{itemize} 

 \end{definition}

\begin{definition}[{Well Formed \LBS\ Program}]
  A \LBS\ {\em program $
  \AC{(D, E, P)}
  $ is  well formed} iff

\begin{tabular}{c}
  \AC{$\WTProc{\defs(E)}{D}$}\\ 
  and   \\   
  $\WTProc{\defs(E),\defs(D)}{P}$
\end{tabular}
\end{definition}

The  {\em big-step operational semantics} of  the expression language
is presented in Fig.\ \ref{fig.opSemExp}. The statement
$\ev{\Pexp}{\val}$ means that the evaluation of $\Pexp$ produces the value $\val$. The rules are {standard}, just notice that selection of the
$i$-th component of a tuple is successful only when the value of the expression to which it is applied has at least $i$ components.
%
We conjecture that evaluation of well typed expressions 
not containing free variables or the metavariable $\this$,
produces a value of the same \AC{type} as the one of the original expression.
\begin{figure}[t]
{
    \begin{mathpar}
       \NamedRule{Exp.ch}{}
      {\ev{a}{a} } 
\and         
       \NamedRule{Exp.const}{}
      {\ev{\const}{\const} } 
\and         
       \NamedRule{Exp.tuple}{\ev{\Pexp_1}{\val_1}\cdots\ev{\Pexp_n}{\val_n}}
      {\ev{\Pexp_1,\ldots,\Pexp_n}{\val_1,\ldots,\val_n} } 
 \and
       \NamedRule{Exp.()}{}
      {\ev{\emptyTuple}{\emptyTuple} }    
\and         
       \NamedRule{Exp.sel}{\ev{\Pexp}{\val_1,\ldots,\val_n}\quad 1\leq i\leq n}
      {\ev{\proj{\Pexp}{i}}{\val_i} } 
 \and         
       \NamedRule{Exp.op}{\ev{\Pexp}{\val}\quad \op(\val)=\val'}
      {\ev{\op(\Pexp)}{\val'} }      
 \end{mathpar}    
    } 
\caption{Operational semantics of expressions}
 \label{fig.opSemExp}
\end{figure}

We now  define  distance between entities, run-time configurations,  structural
equivalence, and  the reduction relation, $\labRed{}$.
\begin{definition}[Distance Between Located Entities]
  We call $\locP{X(\val)}{\sigma}{p}$ a {\em located entity}. If
  $\sigma$ is the shape of $X$, and $\sigma'$ the shape of $Y$,  we define
  \begin{itemize}
  \item $\Points(p,X)=\{p+ q\;|\;q\in\sigma\}$ to be the {\em set of
      points of $X$ positioned at $p$}, and
  \item $\dis{\locP{X(\val)}{\sigma}{p}} {
      \locP{Y(\val')}{\sigma'}{p'}}$ for {\em the distance between two
      located entities}, as the minimum of the set
    $\{d(p_1,p_2)\;|\;p_1\in\Points(p,\sigma)\,\wedge\,p_2\in\Points(p',\sigma')\}$,
    where $d(p_1,p_2)$ is the euclidean distance between the points
    $p_1$ and $p_2$.
  \end{itemize}
\end{definition}

\begin{figure}
{\footnotesize
    \begin{mathpar}
    \NamedRule{S.Loc}
       { P \equiv Q}
     { \locP{P}{\sigma}{p} \equiv \locP{Q}{\sigma}{p} }
     \and 
        \NamedRule{S.Loc.Par}{}
      { \locP{P}{\sigma}{p} \parop \locP{Q}{\sigma}{p} \equiv \locP{P\parop Q}{\sigma}{p} }    
     \and
     \NamedRule{S.Loc.Nu}
       {\ev{(\Pexp[p/\this])}{\rt}\quad\ev{(\Pexp'[p/\this])}{\rd} }
       { (\nu a@\rt,\rd{:}\chType{\ty}).\locP{P}{\sigma}{p}\equiv \locP{(\nu a@\Pexp,\Pexp'{:}\chType{\ty}).P}{\sigma}{p}}    
     \and
   \NamedRule{S.Nu.Com}
       { a\not=b}
       { (\nu a@\rt,\rd{:}\chType{\ty}).(\nu  b@\rt',\rd'{:}\chType{\ty'}).A\equiv 
       (\nu b@\rt',\rd'{:}\chType{\ty'}).(\nu  a@\rt,\rd{:}\chType{\ty}).A}
           \and
   \NamedRule{S.Nu.Abs}
       { }
       { (\nu a@\rt,\rd{:}\chType{\ty}).(\nu  a@\rt',\rd'{:}\chType{\ty'}).A\equiv 
       (\nu a@\rt',\rd'{:}\chType{\ty'}).A}
       \and
       \NamedRule{S.Nu.Par}
    { a\not\in\fns(B)}
     { ((\nu a@\rt,\rd{:}\chType{\ty}).A)\parop B\equiv (\nu a@\rt,\rd{:}\chType{\ty}).(A\parop B)}   
    \end{mathpar}}
  \caption{Structural Equivalence}
  \label{fig.pi.struct}
\end{figure}

\begin{definition}[Spatial Configuration]
  {\em Spatial configurations}, denoted by $A$, $B$, $\ldots$ are
  defined as:
  \[
  A,B \syntaxdef \locP{P}{\sigma}{p} \ou A \parop B \ou \ \new{a@ \rt,
    \rd{:}\chType{\ty}} A \ou \locP{ X(\val) } {\sigma} {p}
  \]
  where $P$ is closed.
\end{definition}
The spatial configuration $\locP{P}{\sigma}{p}$ indicates an entity
that has its barycentre at $p$ and whose behavior is described by the
process $P$, and $\locP{ X(\val) } {\sigma} {p}$ denotes the entity
whose behavior is described by the definition of $X$ and has its
barycentre at $p$. This is different from $\locP{ X(\Pexp)_{\Pexp'} }
{\sigma} {p}$, which represents the entity $X$ evolved from an
unspecified entity {originally} positioned at $p$. The position and the actual
parameter of $X$ will be given by the evaluation of the expressions
$\Pexp'$ and $\Pexp$ respectively, in which the metavariable $\this$ is substituted
by $p$, see function $\place$ below, \AC{which evaluates the locations of
entities.} \AC{This will change
when we add random translation and scaling.}


\AC{
For instance,
in our example from Fig.\ \ref{fig:MicroBioScapeSpatial},
}
the suffix \texttt{this} in the expression \verb!MTRight(y)_this! of
the definition of \texttt{MTPart}, means that the barycentre of a new
instance of \texttt{MTRight} will be the original barycentre of
\texttt{MTPart}.  Another example in the same definition is
\verb!MTLeft(x)_glue(this,u)!, \AC{where \texttt{glue} is an operator
applied to the pair \texttt{(this,u)}, and its value determines the
barycentre of the new instance of \texttt{MTLeft}.
}


The structural equivalence on configurations is defined 
in Fig.\ \ref{fig.pi.struct}, where we omit the rules for associativity and commutativity of $\parop$ and
$+$ and reflexivity, symmetry and transitivity of ${\equiv}$. Parallel composition  has
neutral element $\locP{0}{\sigma}{p}$ for any $p$.
Rule $\la{S.Loc}$ uses
the standard structural equivalence of pi-calculus processes.
In rule $\la{S.Loc.Par}$ the point $p$ is distributed on the two processes saying that both processes
will be located at position $p$. The rest of the rules deal with channel name restriction, and allow us to
bring all the restrictions outside the process, renaming if needed. 
Rule $\la{S.Loc.Nu}$ moves the restriction inside  process located at $p$, 
evaluating the 
expressions for the rate and radius of the channel after the substitution of $p$ for the
metavariable $\this$. Therefore, rate and radius could depend on the location of the
process.

{In the following, the notation $\nu_i$ ($i\geq 0$) is an abbreviation for $\nu a_i@\rt_i, \rd_i:\chType{\ty_i}$.}

\begin{definition}\label{def:canonical}
\begin{itemize}
\item A spatial configuration  {\em $A$ is pre-canonical} if it is of the form:
\[
\nu_1.\ldots.\nu_m.\{ X_1(\Pexp_1)_{\Pexp'_1}\}_{p_1}\parop\cdots\parop\{ X_n(\Pexp_n)_{\Pexp'_n}\}_{p_n}
\]
\item The function
$\place$ is defined as follows:
\begin{enumerate}
\item $\place(\emptyset)=\emptyset$
\item \label{item}$\place(\{ X(\Pexp)_{\Pexp'}\}_{p}\parop A)=\{ X(\val)\}_{p'}\parop\place(A)\}$, where $\ev{\Pexp[p/\this]}{\val}$, and $\ev{\Pexp'[p/\this]}{p'}$
\item {$\place( (\nu a@\rt,\rd{:}\chType{\ty}).\ A) =  (\nu a@\rt,\rd{:}\chType{\ty}).\ \place(A)$}
\end{enumerate}
\item A spatial configuration  {\em $A$ is canonical} if it is of the form:
\[\nu_1.\ldots.\nu_m.\{ X_1(\val_1)\}_{p_1}\parop\cdots\parop\{ X_n(\val_n)\}_{p_n}
\]
\end{itemize}
\end{definition}

The structural equivalence of Fig.~\ref{fig.pi.struct}, allows us to find for 
any $B$, a pre-canonical  $A$ such that $A\equiv B$. 
{The
  function $\place$ evaluates the argument and location of the entity instances
  in the pre-canonical configuration, transforming it into its
  corresponding  canonical configuration. In a canonical configuration all the entities are located.
  {Note that, in the evaluation of both
  $\Pexp$ and  $\Pexp'$ the metavariable $\this$ denotes $p$, the barycentre of the
  entity of which $X$ is the evolution. }
  }

A canonical configuration is {\em  space consistent}, if all its entities are
contained in their respective movement space, and, furthermore, there
are no overlapping entities. The {\it space consistency predicate, {$\OK$}}, is defined as follows. 

\begin{definition}\label{d:OK}
Let $A$ be the canonical configuration $\nu_1.\ldots.\nu_m.\{ X_1(\val_1)\}_{p_1}\parop\cdots\parop\{ X_n(\val_n)\}_{p_n}$. 
$A$ {\em is \OK} if:
\begin{itemize}
\item for all $i$, $1\leq i\leq n$, we have that  $ \Points(p_i,X_i)\subseteq\xi_i$, and
\item  for all $i$, $j$, $1\leq i\not=j\leq n$, we have that $ \Points(p_i,X_i)\cap \Points(p_j,X_j)=\emptyset$. 
\end{itemize}
\end{definition}
{The operational semantics of {\LBS} is given in  Fig.\  \ref{fig.rednorate},
by the reduction relation $\labRed{}$ on {\em run-time
  configurations} of the form
$E\these \{ X_1(\val_1)\}_{p_1}\parop\cdots\parop\{ X_n(\val_n)\}_{p_n}$, where
all the free channel names of $\{ X_1(\val_1)\}_{p_1}\parop\cdots\parop\{ X_n(\val_n)\}_{p_n}$ are
in the domain of $E$}. 
We denote the reflexive and transitive closure of $\labRed{}$ with
$\labRed{}^{\ast}$.
The reduction $\labRed{}$ is defined by the rule $\labP{Par}$.
This rule uses the {\em auxiliary reductions} $\preRedSt{}$  and $\preRedMv{}$.
The spatial configuration $B$ to which $A$ reduces (${E\these A
  \labPreRed{{\tt l}} B} $) {may not be a pre-canonical
configuration, {so, in order 
to produce a correct canonical configuration,
we consider } a pre-canonical configuration, $\nu_1.\ldots.\nu_m.D'$, structurally equivalent to $B$, 
and  then use the function $\place$ to transform it into  a canonical
configuration $D$. }
In the configuration resulting from the reduction, all the channel definitions corresponding to
the restrictions $\nu_1.\ldots.\nu_m$ are moved into the channel environment.
In so doing, we assume renaming of the names in the restriction to avoid clashes  with channel names already in the domain of $E$. 
{In this rule, we also check that the configuration produced is space
consisten, with $D\parop C \ \OK$.} {The rule $\labP{Par}$ cannot be applied, when} there is no auxiliary rule that
can yield a space consistent configuration. 
{The selection of one of the
  choices depends not only on the available interactions with other
  processes, but also on the available space.}
Therefore, the {\em
  evolution of systems in {\LBS} {preserves} space consistency.}

The rules of the auxiliary reductions involve entities, $X(v)$, and entities evolve according to one of the choices  in their
definitions in $D$. In the rules $\labP{Delay}$, $\labP{Com}$, and $\labP{Move}$, there is no  check of 
whether the entities of the resulting process overlap or whether they are contained in their confinement
space. These checks are done, as previously said,  in the
reduction rule $\labP{Par}$. 

In the two stochastic rules, $\labP{Delay}$, and $\labP{Com}$,
$\texttt{r}$ is the rate of the synchronization that determines probability
and duration of the reduction.
Rule $\labP{Delay}$ makes the entity $X$ evolve into the process $P$
with a stochastic rate $\rt$, which is the result of the evaluation of the
expression $\Pexp$ after the substitution of $p$ by $\this$. {Consequently}, the rate may depend on
the position in space of the entity {and its actual parameter}. In rule $\labP{Com}$ the 
entity $X(\val_x)$ sends on channel $a$ the value $\val_a$ to
the entity $Y(\val_y)$, and evolves into process $P$ located at $p_x$. The entity
$Y(\val_y)$ receives $\val_a$ and evolves into $Q$, in which
$\val_a$ substitutes the variable $z$, and it is located at $p_y$. This communication
happens on the common channel $a$, if the located entities
$\{X(\val_x)\}_{p_x}$ and $\{Y(\val_y)\}_{p_y}$ are close enough.
In particular, to interact on channel $a@\texttt{r},\texttt{rad}$, { it must
be the case that} $\dis{\locP{X(\val)}{\sigma}{p}} {
  \locP{Y(\val')}{\sigma'}{p'}}\leq{\tt rad}$. For instance,
${\tt rad}=0$ means that the two entities must be in contact to react.

The non-stochastic rule $\labP{Move}$ defines movement. In this rule,
$\transl(\omega)$ returns {\em a random point whose distance from
  $\Po{0}{0}{0}$ is $\omega$}, and the located entity is moved
randomly a distance $\omega$ from its original position.  This prefix
\verb!mov! says that the entity is subject to brownian motion.
\begin{figure}[tp]
{
  \begin{mathpar}
    \NamedRule{Par} { {E\these A \labPreRed{{\tt l}} B} \quad\quad
     B\equiv\nu_1.\ldots.\nu_m.D'\ \
     \mbox{pre-canonical}\quad\quad\place(D')=D\quad\quad D\parop C\ \
     \OK \quad {{\tt l} \in \{\rt , \mbox{{\tt mv}} \} }} { E\these
      A\parop C \labRed{} E,a_1@\rt_1, \rd_1{:}\chType{\ty_1},\ldots,a_m@\rt_m, \rd_m{:}\chType{\ty_m}\these D\parop C }
 \and
     \NamedRule{Delay}
        {X(x)= ({\nu_1.\ldots.\nu_n.}delay@\Pexp.P\;[+\;M])^{\xi,\omega,\sigma}\in D\quad\ev{\Pexp[p/\this,\val/x]}{\rt} }
        {E\these \locP{X(\val)} {\sigma}{p}\preRedStD{\rt}  \locP{{\nu_1.\ldots.\nu_n.}P[\val/x]}{\sigma}{{p}}}
  \and
  \NamedRule{Com}
        {\begin{array}{l}
         X(x)={\nu_1.\ldots.\nu_n.}M_x^{\xi,\omega,\sigma}\in D\quad M_x[\val_x/x]=(!a(\Pexp_a).P\;[+\;M])\quad\ev{\Pexp_a[p_x/\this]}{\val_a}\\
         Y(y)= {\nu'_1.\ldots.\nu'_m.}M_y^{\xi',\omega',\sigma'}\in D\quad M_y[\val_y/y]=(?a(z).Q\;[+\;N])\\
         \dis{\locP{X(\val_x)} {\sigma}{p_x} }{ \locP{Y(\val_y)}{\sigma'}{p_y} } \leq\rd
        \end{array}
       }
         {E,   a@\rt, \rd:\chType{\ty}\these  \locP{X(\val_x)}{\sigma}{p_x} \parop
          \locP{Y(\val_y)}{\sigma'}{p_y}
          \preRedSt {}  \locP{ {\nu_1.\ldots.\nu_n.}P} {\sigma} {p_x} \parop
          \locP{{\nu'_1.\ldots.\nu'_m.}Q[\val_a/z]} {\sigma'}{p_y} }
   \and
  \NamedRule{Move}
      { p'=p+\transl(\omega)  \quad
        X(x)= ({\nu_1.\ldots.\nu_n.}\mov.P \;[+\;M])^{\xi,\omega,\sigma}\in D}
      { E\these\locP{X(\val)}{\sigma'}{p} \preRedMv{}
        \locP{{\nu_1.\ldots.\nu_n.}P[\val/x]}{\sigma'}{p'} }
    \end{mathpar}
    }
  \caption{{Reduction Relation}}
  \label{fig.rednorate}
\end{figure}

{We conjecture that 
if $(D,E,\locP{X_1(\val_1)}{}{p_1}\parop\cdots\parop
\locP{X_n(\val_n)}{}{p_n})$ is a well formed \LBS\ program, then for all $E'$ and $A$ such that
$\defs(E), \defs(D)\these \locP{X_1(\val_1)}{}{p_1}\parop\cdots\parop
\locP{X_n(\val_n)}{}{p_n}\labRed{}^{\ast}E'\these A$, we have that
$\WTProc{E'} {A}$. }

\section{Random Translation and Scaling}\label{sec:RandomScale}
\paragraph{Random Translation} Consider the case of a bacterium that secretes a hydronium ion
(\verb+HIon+). The language extension discussed so far will allow us
to describe where to locate the \verb+HIon+, but it will be at a
specified location with respect to the position of the
bacterium. Instead we would like to be able to say that it should be
at a given distance, but in a random direction. To this end, we annotate entity instances
with expressions evaluating to pairs, whose first component is, as
before, a translation point, and the second \AC{component}, a number which specifies a distance from which
we generate a random point, as in the rule \labP{Move} of
Fig~\ref{fig.rednorate}. In the fragment of code in
Fig.~\ref{fig:bac}(a),
the barycentre of the instances of \verb!Bac! will be in the position of the \verb!Bac!
they evolve from. On the other hand, the barycentre of the instances of \verb!HIon! 
will be in a random position that is at a distance equal to the sum of the radii of the bacterium and
the ion, from the barycentre of the \verb!Bac! it evolves from.

\begin{figure} [ht]
\centering
  {
\begin{verbatim}
Bac()@_,_,_ =                         
do mov.Bac()_(this,0) 
or delay@0.005.(Bac()_(this,0) | HIon()_(this,rB+rH)) 
or ...              
\end{verbatim}  
(a)
\begin{verbatim}
Bac()@_,_,_,max-size =                       
do mov.Bac()_((fst(this),0),1.1) 
or delay@0.005.( Bac()_((fst(this),0),1) | HIon()_((fst(this),rB+rH),1) ) 
or delay@0.2.( Bac()_((fst(this),rB),0.5) | Bac()_((fst(this),rB),0.5) )
or .....              
\end{verbatim}
(b)
}
  \caption{(a) Random translation and (b) scaling}
\label{fig:bac}
\end{figure}

{As far as the definition of the syntax  for  this extension, we have to change the typing rule for entity
instances so that the type of the subscript expression, $\Pexp'$, is a pair whose first component 
has the type of a point (giving the deterministic component of the
translation) and the second component is 
a floating point (giving the random component of the translation). The new rule is $\labP{Ty.inst.R}$ of Fig. \ref{fig.typingRS}.}
{Notice that, up  until now, given an entity instance.
$X(\Pexp)_{\Pexp'}$, the metavariable $\this$ and $\Pexp'$ had the same
type. However, this is} {no longer the case, since,
even though the expression $\Pexp'$ has type $(\floatType\ast\floatType\ast\floatType)\ast\floatType$, the metavariable $\this$  still has type 
$\floatType\ast\floatType\ast\floatType$.}

\begin{figure}[ht]
{
  \begin{mathpar}
     \NamedRule{Ty.inst.R}{X{:}\ty\in\tyEnv\quad\WTExp{\tyEnv}{\Pexp}{\ty}\quad\WTExp{\tyEnv}{\Pexp'}{(\floatType\ast\floatType\ast\floatType)\ast\floatType} }
    {\WTProc{\tyEnv} { X(\Pexp)_{\Pexp'}}}    
\and      
    \NamedRule{Ty.inst.RS}{X{:}\ty\in\tyEnv\quad\WTExp{\tyEnv}{\Pexp}{\ty}\quad\WTExp{\tyEnv}{\Pexp'}{((\floatType\ast\floatType\ast\floatType)\ast\floatType)\ast\floatType} }
    {\WTProc{\tyEnv} { X(\Pexp)_{\Pexp'}}}  \and 
        \NamedRule{Ty.this.RS}{}
      {\WTExp{\tyEnv}{\this}{(\floatType\ast\floatType\ast\floatType)\ast\floatType} }  
    \end{mathpar}
}
 \caption{Typing rule for entity instance and $\this$ for random translation and scaling}
  \label{fig.typingRS}
\end{figure}

Regarding the semantics, the located entities are still
annotated with a point, however,
we have to change the definition of entity placing into space, since
now, the evaluation of $\Pexp'$ (the subscript of the entity instance) 
produces a pair, whose first component is a point, giving the deterministic translation, and 
the second \AC{component is} a floating point giving the length of the random translation. To this extent,
clause (\ref{item}) of function $\place$ in Definition~\ref{def:canonical} is modified as follows:
\[
\place(\{ X(\Pexp)_{\Pexp'}\}_{p}\parop A)=\{
  X(\val)\}_{p'+\transl(\const)}\parop\place(A)
  \]
where $\ev{\Pexp[p/\this]}{\val}$ and
$\ev{\Pexp'[p/\this]}{(p',\const)}$. {Notice that while $\Pexp'$ evaluates to a
pair, and it is specified by the programmer, $p'+\transl(\const)$ evaluates to a
point (a triple of floating points).}

\paragraph{Scaling} We now consider the shape of entities. As it is now, we have a specific shape and always the same dimension. In
order to represent a change in scale, a new entity with a smaller or
bigger shape would have to be defined. Alternatively, 
we would like to
be able to change the size of the entity using scaling directives.
For instance, consider adding to the previous example of the bacterium the fact that 
bacteria grow and divide, and that movement is associated with a growth
of 10\%.  
Accordingly, we add this
behavior in Fig.~\ref{fig:bac}(b), by specifying that the bacterium may spontaneously
divide into two bacteria of half the size of the original one (0.5), and
moved apart in random directions  a distance equal to the radius of
the shape of the original bacterium (\texttt{rB}). Moreover, 
movement is associated with a growth
of 10\% (1.1).

{As far as the syntax of the language is concerned,  instances are
  annotated with an expression, $\Pexp'$, evaluating {to a pair $((p,\const),s)$, whose first component 
  is also a pair, $(p,\const)$, which specifies, as in the random translation extension, the
  deterministic and random components of the translation.}
\AC{The second component of the evaluation of  $\Pexp'$,  is $s$, the scaling factor for the shape.}
A located entity is now characterized by having a location for its barycentre,
and a scaling factor affecting its shape. Consequently, the metavariable $\this$, 
denotes a \AC{ pair: point and scaling factor.}
In Fig.~\ref{fig.typingRS}, we give the new typing rules: {$\labP{Ty.inst.RS}$} for entity instances, and
{$\labP{Ty.this.RS}$}, for $\this$. {We use $\fst$ and $\snd$ to access the first and second 
component of a pair.}

Going back to the example of Fig.~\ref{fig:bac}(b),
  \verb+mov.Bac()_((fst(this),0),1.1)+  {specifies that the barycentre of this
  instance of \texttt{Bac} after \texttt{mov}, will be the same as the one of the 
\texttt{Bac} it evolved from, since  \texttt{fst(this)} is the barycentre of the generating \texttt{Bac} and the random
translation is of length $0$. The scaling $1.1$ gives a 10\% increase in size with respect to the generating \texttt{Bac}.
Additionally, in the definition of the  entity \texttt{Bac}, we fix a
growth limit with \texttt{max-size}. Finally the original move rule
will generate the position of the located entity according to the
diffusion rate of \texttt{Bac}.}

The second line of the definition, specifying the secretion of the \texttt{HIon} is as for the example in Fig.~\ref{fig:bac}(a). 
Just note that, the barycentre of the instance of \texttt{Bac} is specified by \texttt{(fst(this),0)}, i.e., it is the same as 
the one of the generating \texttt{Bac},
and the one of \texttt{HIon} is \texttt{(fst(this),rB+rH)},  i.e., at a distance \texttt{rB+rH} from  
the one of the generating \texttt{Bac}.

Finally, \verb+delay@0.2.(Bac()_((fst(this),rB),0.5)|Bac()_((fst(this),rB),0.5))+  {specifies that 
the two instances of the daughter bacteria are positioned at a distance \texttt{rB} from the generating bacterium,
\texttt{(fst(this),rB)}, and they are half its size, $0.5$. }

{
Spatial configurations now record the scaling factor $s$, in addition to the
barycentre of the shape of the entity, and are defined as follows.}
\[   A,B   \syntaxdef \locP{P}{\sigma}{(p,s)} \ou A \parop B \ou \
   \new{a@ \texttt{r}, \texttt{rad}} A \ou \locP{ X(\val) } {\sigma} {(p,s)}
\]
The structural equivalence rules of Fig.~\ref{fig.pi.struct} are modified by replacing $(p,s)$ 
for $p$, in rules $\la{S.Loc}$, $\la{S.Loc.Par}$, and
$\la{S.Loc.Nu}$. We also  modify clause (\ref{item}) of function $\place$ in Definition~\ref{def:canonical}, as follows: 
\[ 
\place(\{ X(\Pexp)_{\Pexp'}\}_{(p,s)}\parop A)=\{ X(\val)\}_{(p'+\transl(\const_r\times s),s\times\const_s)}
\] 
where $\ev{\Pexp[(p,s)/\this]}{\val}$, and $\ev{(\Pexp'[(p,s)/\this])}{((p',\const_r),\const_s)}$.
As we can see, both the length of the random translation and 
 the scaling factor of the located entity $X$ are obtained by multiplying the result of
 the evaluation of the expression $\Pexp'$ by the  scaling factor $s$ of the entity
from which $X$ evolved.  

{The auxiliary reduction relations are obtained from the rules
  Fig.~\ref{fig.rednorate}, by replacing
$(p,s)$ for $p$ in  rule $\la{Delay}$,  $(p_x,s_x)$ for $p_x$, and $(p_y,s_y)$ for $p_y$, 
in rule $\la{Com}$. Moreover, since
scaling affects the dimension of the shape of entities, the definition of the distance between 
entities will have to take  into account this fact. In particular,
$\dis{\locP{X(\_)} {\sigma}{(p_x,s_x)} }{
  \locP{Y(\_)}{\sigma'}{(p_y,s_y)} }$ is the minimum of the following set.
 \[
 \{d(p_1,p_2)\;|\;p_1\in\Points(p_x,\Scale(s_x,X))\,\wedge\,p_2\in\Points(p_y,\Scale(s_y,Y))\},
 \] 
 where $\Scale(s,X)=\{s\times p\;|\;p\in\sigma\}$. 

 Finally, in rule
$\labP{Move}$ we have to scale the random quantity added to the translated point since
this quantity refers to the initial (standard) dimension of entity
$X$.  The new rule 
 $\labP{Move.S}$ is shown in Fig.\ \ref{fig:newrules}.
 }
\begin{figure} 
{
  \begin{mathpar}
  \NamedRule{Move.S} {
      X(x)= ({\nu_1.\ldots.\nu_n.}\mov.P \;[+\;M])^{\xi,\omega,\sigma}\in D
} 
{
    \locP{X(\val)}{\sigma'}{(p,s)} \preRedMv{}
    \locP{{\nu_1.\ldots.\nu_n.}P[\val/x]}{\sigma'}{(p+\transl(s\times\omega) ,s)} }
    \end{mathpar}
}
 \caption{Modified rule $\labP{Move}$ for random translation and scaling}
  \label{fig:newrules}
\end{figure}

Since scaling affects the dimension of the shape of entities, and therefore the space occupied  by them,
 {the definition of space consistent configuration (Definition
   \ref{d:OK}) }  is modified as follows.
\begin{definition}\label{d:OK-RS}
Let $A$ be the canonical configuration $\nu_1.\ldots.\nu_m.\{ X_1(\val_1)\}_{(p_1,s_1)}\parop\cdots\parop\{ X_n(\val_n)\}_{(p_n,s_n)}$.  $A$ {\em is \OK} if:
\begin{itemize}
\item for all $i$, $1\leq i\leq n$, we have that  $ \Points(p_i,\Scale(s_i,X_i))\subseteq\xi_i$, 
\item for all $i$, $1\leq i\leq n$, we have that  $s_i\leq\mu_i$, and
\item  for all $i$, $j$, $1\leq i\not=j\leq n$ we have that $ \Points(p_i,\Scale(s_i,X_i))\cap \Points(p_j,\Scale(s_j,X_j))=\emptyset$. 
\end{itemize}
\end{definition}
{We  still conjecture that, starting from a space consistent initial
configuration we get subsequent space consistent configurations.}
        
\section{Related Work}
In this paper we define \LBS, a stochastic pi-calculus in 3D-space
with programmable locations. It is an  extension of BioScape\cite{BioScape:ENTCS},
of which the authors have also provided a parallel semantics \cite{PBioScape:Mecbic}. 
We have introduced the language with its type system  and the operational semantics.
The position and size of the entities can be programmed,
and the operational semantics enforces the constraint that during the 
evolution of the system, entities are confined in their containing space and
do not overlap.

\LBS, like BioScape, \AC{is a stochastic process algebras}. As an alternative to
models built around sets of ordinary differential equations (ODEs),
process algebras are formal languages where multiple objects with
different behavioral attributes can interact with each other and
dynamically influence overall systems development. Process 
algebras have been originally introduced for the description of complex
reactive processing systems. The description of a system is modular, with sub-components 
interacting through shared channels. This is similar to the structure 
of biological systems where species can be seen as processes, and their
interaction with other species is described by synchronization on channels.
In stochastic process algebras, synchronization happens at a stochastic rate.
This reflects more closely the behavior of biological system. Some
stochastic process algebras which have been proposed, are PEPA
\cite{Hillston05} and  EMPA \cite{BernardoG98}. With name passing stochastic process algebras, 
such as stochastic pi-calculus \cite{Priami95}, information or data can be exchanged
on communication channels. By using the tool SPiM (Stochastic Pi
Machine) \cite{Phillips07}, computer simulations can be run, and the change 
in time of the biological species is displayed. A number of biological systems have 
been modeled with the use of stochastic pi-calculus \cite{Priami01,CardelliGTP09,Baoetal2010,DBLP:conf/scsc/SharmaC13}. 

One of the motivations for the introduction of \LBS {} is the desire to model
biological systems in which position of entities in space could be used to 
determine their behavior. A limited notion of space 
is incorporated in BioAmbients \cite{BiAM} and BioPepa \cite{Biopepa09}.  
Geometric capabilities are 
present in \AC{a spatial}  extension of the pi-calculus \cite{spacePi}, Shape Calculus \cite{DBLP:journals/cuza/BartocciCBMT10}, and CCS-like timed
calculus with an associated simulating tool \cite{BioShape}.
However, both \cite{spacePi} and \cite{DBLP:journals/cuza/BartocciCBMT10} lack stochasticity. As already mentioned 
in the introduction, the calculus that is closer to {\LBS} is \ThreePi\cite{DBLP:journals/tcs/CardelliG12}, 
a geometric process algebra in which the processes are equipped with affine transformations. 
There are two main differences between {\LBS} and \ThreePi.
First, in {\LBS} we do not consider affine transformations, but just 
a uniform scaling in all directions maintaining the 
barycentre of the entity in its original position, and in 
addition to standard translation also a random translation. \AC{Neither our scaling nor 
our random translation are  affine transformations}. Second, and most important, 
is the fact that \ThreePi{} is a low level language that gives
absolute control of spatial attributes to the programmer, while in 
{\LBS} the programmer specifies species at a higher level,
and it has been designed to program biological and biomaterial
processes and their interactions.
In \ThreePi, diffusion and confinement, have to be explicitly controlled 
by the programmer in terms of 
the low level abstraction provided by affine transformations.
In \cite{CardelliActin09} an extension of pi-calculus for displaying 
geometric information is introduced. However, this is \AC{a rather } \textit{ad hoc} 
extension motivated
by the description of the biological processes to model actin polymerization.

\section{Future Work}
In collaboration with materials scientist Matthew Libera, from
the Stevens Institute of Technology, we are working on the computationally assisted development of
antibacterial surfaces \cite{Sharma2013}. 
{Traditionally biomaterials development consists of designing a surface
and testing its properties experimentally. This trial-and-error
approach is limited, because of the resources and time needed to sample
a representative number of configurations in a combinatorially complex
scenario.}  In many cases the design is also aided by computational
models tailored to a specific application. In these cases, there have
been successful attempts to identify biomaterials with optimal
properties \cite{CAD,MARC:MARC200300193,Zygourakis1996}. However,
developing such dedicated software frameworks is time consuming, and
small modifications in the understanding of the application can lead
to significant and time consuming software changes.

Our proposal consists of  designing  antibacterial biomaterials
from first principles. Using the antibacterial
effect of individual components, we will computationally design
optimally antibacterial surfaces, which simultaneously promote the
growth of healthy tissue. 
Our model will stochastically assemble
surface blocks whose connectivity will be determined by their
antibacterial properties, as well as their ability to encourage tissue
growth, in the same way a child assembles building blocks. These
designed  surfaces will then be tested in virtual
experiments in the same platform. 
In order to test these surfaces we will use \LBS, where surfaces will
be described by a collection of located entities generated by the
surface design process.

The emerging surface patterns with maximal antibacterial effect will
be used to design tiling patterns, which will motivate the design of
new biomaterials that will then be tested in wet lab experiments.

\section{Acknowledgements}

We are grateful to Philip Leopold for introducing us to the
fascinating world of intracellular transport. We also thank
Mariangiola Dezani for illuminating discussions and comments on earlier
drafts. This article was also improved due to the helpful suggestions
of the anonymous referees of DCM 2013. Vishakha Sharma acknowledged the generous support from the Stevens Center for Complex
Systems and Enterprises.

  \bibliographystyle{eptcs} 
  \bibliography{TPrefs}

\end{document}